\documentclass[aps,prb,twocolumn,groupedaddress,showpacs,floatfix,superscriptaddress,longbibliography,10pt]{revtex4-2}
\usepackage{float}
\usepackage[utf8]{inputenc}
\usepackage[T1]{fontenc}
\usepackage{mathtools,amssymb,graphicx,units,bm}
\usepackage{soul} 
\usepackage[dvipsnames]{xcolor}
\usepackage[normalem]{ulem}
\usepackage{bbold} 
\usepackage{dcolumn}
\usepackage[version=4]{mhchem}
\usepackage{physics}%
\usepackage{silence}
\usepackage[plainpages=false,pdfpagelabels,colorlinks=true,linkcolor=red,urlcolor=PineGreen,citecolor=PineGreen,pdftitle={NHSSH},pdfauthor={SMGMA},pdfdisplaydoctitle=true,pdfduplex=DuplexFlipLongEdge]{hyperref}
\usepackage{orcidlink}

\newcommand{\prlsection}[1]{\noindent\textit{#1.---}\hspace{0.25em}}

\newcommand{\psilk}{\psi_{\rm L}}
\newcommand{\psirk}{\psi_{\rm R}}
\newcommand{\dn}{\delta_n}

\mathchardef\mhyphen="2D 

\begin{document}

\title{Enhancement of charge correlations and real-space topological marker on an interacting non-Hermitian Su-Schrieffer-Heeger model}

\author{Sebasti\~ao dos A. Sousa-Júnior\,\orcidlink{0000-0002-4266-3780}}
\email{sebastiao.a.s.junior@gmail.com}
\affiliation{Department of Physics, University of Houston, Houston, Texas 77204, USA}
\author{Pedro B. Melo\,\orcidlink{0000-0002-3726-761X}}
\affiliation{Departamento de F\'isica, PUC-Rio, 22452-970, Rio de Janeiro RJ, Brazil}
\affiliation{Universit\`a degli Studi di Palermo, Dipartimento di Fisica e Chimica - Emilio Segr\`e, via Archirafi 36, I-90123 Palermo, Italy}
\author{Rubem~Mondaini\,\orcidlink{0000-0001-8005-2297}}
\affiliation{Department of Physics, University of Houston, Houston, Texas 77204, USA}
\affiliation{Texas Center for Superconductivity, University of Houston, Houston, Texas 77204, USA}
\author{Arnob Kumar Ghosh\,\orcidlink{0000-0003-0990-8341}}
\affiliation{Department of Physics and Astronomy, Uppsala University, Box 524, 75120 Uppsala, Sweden}
\author{Rodrigo Arouca\,\orcidlink{0000-0003-4214-1437}}
\affiliation{Brazilian Center For Research in Physics, Rua Doutor Xavier Sigaud 150, Rio de Janeiro, 22290-180, Brazil}


\begin{abstract}

We investigate the interacting non-Hermitian Su-Schrieffer-Heeger (SSH) model, focusing on the interplay between topology and charge ordering. Using a real-space topological marker, charge correlations, and the complex many-body spectrum, we map out the phase diagram under periodic and open boundary conditions. We show that the topological marker remains a robust diagnostic of non-Hermitian topological phases in the presence of interactions and consistently signals their breakdown at the onset of a charge density wave (CDW). We further demonstrate that non-Hermiticity enhances interaction effects: While moderate changes occur under periodic boundary conditions, open boundary conditions lead to a pronounced amplification of staggered charge correlations near exceptional points. This enhancement arises from the accumulation of low-energy states near exceptional points, which promotes electronic instabilities and strengthens CDW tendencies.

\end{abstract}
 
\maketitle

\prlsection{Introduction} Topological phases are a rich and dynamic field of research in condensed matter physics characterized by topologically protected boundary modes~\cite{hasan2010colloquium, qi2011topological, BernevigBOOK}. These are characterized by bulk invariants that, under suitable conditions, predict protected boundary modes~\cite{Altland97, Chiu16}, an aspect known as bulk-boundary correspondence~\cite{hasan2010colloquium, qi2011topological, BernevigBOOK}. Such a paradigm directly applies to situations in which an effective single-particle description can classify these modes, with notable examples including topological phases protected by crystalline symmetries~\cite{Fu2011, Slager2013, Po2017} and periodic-in-time evolution operators described by Floquet theory~\cite{Kitagawa10, Kitagawa11, Lindner2011, Ghosh_2024}.

In the past decade, the realm of topological phases has been extended to non-Hermitian Hamiltonians, which are effective descriptions of open quantum systems, at least at short times~\cite{BreuerPetruccione2002, El-Ganainy2018, Kawabata2019, Song20192, Minganti2019, Lieu2020, Ashida2020, DErrico2020, Bergholtz2021, Lee2023, YangPRR2023, GhoshNW2024, salatino2025}. Non-Hermitian Hamiltonians extend the conventional 10-fold Altland-Zirnbauer~(AZ) classification to 38-fold~\cite{bernard2002classification, Gong2018, Kawabata2019}, owing to the ramifications of the symmetry operations. Furthermore, non-Hermitian systems exhibit the non-Hermitian skin effect~\cite{Lee16, FoaTorres18, Lee19}, a density accumulation at a given edge in a quantum system, which has a topological origin by itself~\cite{shen2018, Okuma20, Lin2023}. The skin effect makes the system's spectrum highly sensitive to boundary conditions, thereby modifying the conventional bulk-boundary correspondence \cite{Yao2018SSH, Yao2018, Kunst18, Yokomizo2020}. Moreover, non-Hermitian matrices exhibit exceptional points (EP)~\cite{Berry2004, Heiss_2012, Miri2019}, i.e., degeneracies of their eigenpairs that modify the possible gapless topological phases~\cite{Maliybaev05, shen2018, Zhou18, Carlstrom18, Budich19, Carlstrom19, Bergholtz2021, Yang_2024, Kozii24}.

Although non-Hermitian topological systems have been extensively studied in the last decade, their interacting counterparts are only recently getting more attention~\cite{ZhangNHIntPRB2020, LiuNHIntPRB2020, FaugnoNHIntPRL2022, KawabataNHIntPRB2022, Zhang2022, YoshidaNHIntPRB2023, Wang2023, RoccatiNC2024, Zhong2025, Yi2025}. In general, interactions can completely change the possible topological phases \cite{WenScience}, and approaches such as group cohomology~\cite{Chen13, turner2013}, ground-state degeneracies~\cite{Levin06}, and anomalies~\cite{Arouca2022} are often employed to classify them. 

For numerical many-body methods formulated in real space, such as exact diagonalization (ED)~\cite{Melo2023} and density matrix renormalization group (DMRG)~\cite{Antao2025, Zhong2025}, local topological markers offer a particularly useful alternative~\cite{Bianco2011, WeiChen2023, Lage25, Sousa25}, since they can diagnose topology without relying on translational invariance. Indeed, in one-dimensional topological systems, for example, the Zak phase~\cite{Zak1989, Berry1984} characterizes the occupied-state topology and serves as a diagnostic of the underlying winding structure. In simulations, this quantity is commonly obtained by introducing twists in the boundary conditions and following the associated Berry phase, a strategy that has been implemented in DMRG calculations~\cite{Zaletel2014, Birnkammer2022} and generalized to topological responses in higher dimensions~\cite{Grushin2015, Shao2021b, He2024}. These approaches, however, generally require repeated calculations over boundary twists, leading to a substantial computational overhead. By contrast, a topological marker can be extracted from a single boundary condition, including open boundaries, and has been developed for noninteracting systems, density matrices, and interacting Hermitian models~\cite{Bianco2011, Song2019, Bardyn2018, Melo2023, Hannukainen2024, sousa26}.
 
\begin{figure}[t]
    \centering
    \includegraphics[width=0.47\textwidth]{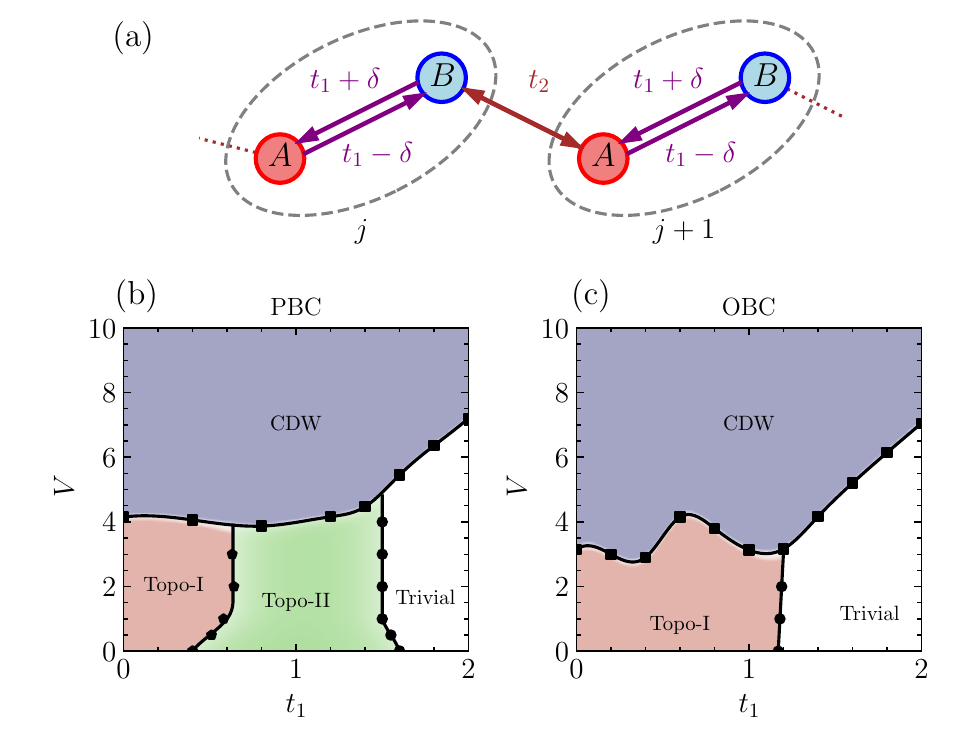}
    \caption{(a) Schematic representation of the hoppings in the non-Hermitian dimerized SSH model, see Eq.~\eqref{eq:Ham_chosen}. Phase diagram of the system for periodic, (b) [open, (c)], boundary conditions. Topo-I and Topo-II stand for the topological Hermitian and non-Hermitian phases, respectively, with unity or half-unity winding number, and the trivial insulator phase with zero winding number. Phase boundaries are determined by finite-size scaling of charge correlations, many-body energy gaps, and the topological marker. Data are shown for $t_2 = 1$, $\delta = 0.6$, computed from left and right eigenvectors.
    }
    \label{fig:lattice}
\end{figure}

Here, we investigate the fate of the topological phases of the one-dimensional (1D) non-Hermitian Su-Schrieffer-Heeger (SSH) model \cite{Yao2018SSH, Yin2018, Arouca2020} in the presence of interactions~\cite{Zhong2025}. We use ED to solve the system and determine the topological phases, using a local topological marker for both periodic boundary conditions (PBC) and open boundary conditions (OBC) --- results are summarized in Fig.~\ref{fig:lattice}. We find a competition between the charge density wave (CDW) phase and two topological phases, topological-I (Topo-I) and topological-II (Topo-II), within PBCs. With OBC, we observe a competition between the Topo-I and CDW phases. In contrast to the Hermitian case \cite{Melo2023},  interactions enlarge the region of the topological phase, as they suppress the non-Hermitian phase more strongly. Moreover, the presence of a higher-order EP with OBCs significantly enhances charge correlations due to the increment of density of states, similarly to what has been shown for topological superconductors in Ref.~\cite{Arouca2023}. 


\prlsection{Model}
The Hamiltonian of the effective non-Hermitian 1D SSH model in the presence of nearest-neighbor interaction and non-reciprocal hopping reads as~\cite{Arouca2020, Melo2023}
\begin{align}
    \hat{\cal H} =  \sum_{j=1}^{N_{\rm cell}} & \Big[ (t_1-\delta) \hat c^\dagger_{j,B}   \hat c_{j,A}^{\phantom{\dagger}} + (t_1+\delta) \hat c^\dagger_{j,A}   \hat c_{j,B}^{\phantom{\dagger}} \nonumber \\
    & + t_2 \hat c^\dagger_{j+1,A}\hat c_{j,B}^{\phantom{\dagger}} + t_2 \hat c^\dagger_{j,B}   \hat c_{j+1,A}^{\phantom{\dagger}}  \nonumber \\
    & + V \hat n_{j,A} \hat n_{j,B} + V \hat n_{j,B} \hat n_{j+1,A} \Big],
     \label{eq:Ham_chosen}
    \end{align}
where the operator $\hat c^\dagger_{j,\alpha}$~($\hat c_{j,\alpha}^{\phantom{\dagger}}$) creates (annihilates) a spinless fermion at sublattice $\alpha~(=A, B)$ of the $j$-th unit cell, $\hat n_{j,\alpha} = \hat c^\dagger_{j,\alpha} \hat c_{j,\alpha}^{\phantom{\dagger}}$ is the number operator, and $N_{\rm cell}$ stands for the number of unit cells. 

The parameters $t_1 \pm \delta$ denote the intra-cell hopping amplitudes connecting the $A$ and $B$ sites within the same unit cell, where $\delta$ introduces a non-reciprocal hopping, and thereby makes the system non-Hermitian. Such an effective non-Hermitian term arises when considering an open system with directional coupling between system and reservoir \cite{Gong2018, Guo2020, Liang2022, Ekman2025Lic}. In the supplementary material~(SM)~\cite{supp}, we derive the effective non-Hermitian Hamiltonian starting from the Lindblad master equation. The terms proportional to $t_2$ represent the inter-cell hopping between sublattices $A$ and $B$. The last two terms describe the nearest-neighbor density-density interactions: $V \hat n_{j,A} \hat n_{j,B}$ accounts for the interaction within each unit cell, while $V \hat n_{j,B} \hat n_{j+1,A}$ represents the interaction between neighboring cells. A schematic representation of our model Hamiltonian [Eq.~\eqref{eq:Ham_chosen}] is illustrated in Fig.~\ref{fig:lattice}(a). We define the total number of sites as $N = 2N_{\rm cell}$ and set $t_2 = 1$ throughout unless otherwise specified. 
\vskip0.05in
\prlsection{Topological marker and structure factor} In interacting systems, the full characterization of topology occurs in sectors containing more than one particle. Here, we consider a half-filled chain with an average occupancy $n \equiv \sum_j \langle \hat n_{j,A} + \hat n_{j,B}\rangle/N_{\rm cell} = 1/2$ particles per site. Thus, a suitable method for probing topology is based on a real-space marker~\cite{Bianco2011, WeiChen2023, Melo2023}. Before moving to the interacting case, we briefly discuss the real-space marker in the absence of interactions ($V=0$).

In a Hermitian system, the real space marker is evaluated by summing over the occupied states (typically those with energy $E < 0$)~\cite{Bianco2011, WeiChen2023}. However, in a non-Hermitian system, the notion of occupied states is no longer well-defined, as the eigenenergies are generally complex. Nevertheless, we can utilize the properties of the eigenstates under chiral symmetry $S$~\cite{Song2019}. In particular, the right eigenstates satisfy: $\hat {\cal H}\ket{\psi_{\rm R}^\nu} = E_\nu \ket{\psi_{\rm R}^\nu}$ and $\hat {\cal H}\ket{\psi_{\rm R}^\mu} = -E_\nu \ket{\psi_{\rm R}^\mu}$; where $\ket{\psi_{\rm R}^\mu}$ being proportional to $\hat {S}\ket{\psi_{\rm R}^\nu}$. Thus, the eigenstates are related by the chiral symmetry, which is also respected by our system. A similar relation holds for the left eigenstates, which satisfy $\hat {\cal H}^\dagger\ket{\psi_{\rm L}^\nu} = E_\nu^* \ket{\psi_{\rm L}^\nu}$ and $\hat {\cal H}^\dagger\ket{\psi_{\rm L}^\mu} = -E_\nu^* \ket{\psi_{\rm L}^\mu}$. Thus, by denoting the pair of chiral-mirrored eigenstates as $\{\nu,\mu\}$, we construct the projector $\hat P$~\cite{Melo2023}, whose matrix elements are given by
\begin{align}
    P_{rr^\prime} = \sum_{\nu\in S} \frac{\bra{\psi_{\rm L}^\nu} \hat c_r^\dagger \hat c_{r^\prime}^{\phantom{\dagger}} \ket{\psi_{\rm R}^\nu}}{\braket{\psi_{\rm L}^\nu}{\psi_{\rm R}^\nu}}\ .
    \label{eq:projector}
\end{align}
In principle, one would expect bi-orthogonality of the eigenpairs, $\braket{\psi_{\rm L}^n}{\psi_{\rm R}^m} = \delta_{n,m}$; however, this is not numerically exact, and thus, we normalize the matrix elements by $\braket{\psi_{\rm L}^\nu}{\psi_{\rm R}^\nu}$.
Defining the complementary operator $\hat Q = \mathbb{1} - \hat P$, the real-space topological marker is expressed as~\cite{Song2019, supp}
\begin{equation}
W = \frac{1}{N_{\rm C}}\mathrm{Tr}_{\rm C}\!\left( \hat S \hat P \hat X \hat Q + \hat S \hat Q \hat X \hat P \right)\ ,
\label{Eq:marker}
\end{equation}
where $\hat X$ is the unit-cell position operator, and $\mathrm{Tr}_{\rm C}$ denotes the trace restricted to the central region of the lattice of length $N_{\rm C} = N/2$. Using $W$, one can compute the phase diagram of the non-interacting system for both OBCs and PBCs, see Sec.~\ref{subsec:NIS} of the SM~\cite{supp}. In addition, we show that computing the topological marker solely from the right eigenvectors, unlike from both $\ket{\psi_{\rm L}^\nu}$ and $\ket{\psi_{\rm R}^\nu}$ as highlighted in Eq.~\eqref{eq:projector}, fails to reproduce the correct phase diagram for this model~\cite{Yin2018, Arouca2020}.

We are now in a position to generalize it to the many-body case. For that, we extract the ground state of the non-Hermitian SSH model using a Krylov-Schur diagonalization algorithm~\cite{Balay, Slepc}. In this approach, we target the eigenpair (left and right) with the smallest real part of the eigenvalue. In cases where multiple eigenvalues share the same real part, the eigenstate with the smallest imaginary part is selected. This eigenpair, denoted by $\ket{\Psi_{\rm L}^0}$ and $\ket{\Psi_{\rm R}^0}$, is referred to as the ground-state eigenpair throughout this work. The many-body energy gap is defined as the difference between the two eigenvalues with the smallest real parts; when degenerate, the ground state is the one with the smallest imaginary part. Utilizing the ground state eigenpair, we compute the  matrix elements of the projector $\hat P$ as
\begin{align}
    P_{rr^\prime}^{\eta \rm R} =  \frac{\bra{\Psi_{\eta}^0} c_r^\dagger c_{r^\prime}^{\phantom{\dagger}} \ket{\Psi_{\rm R}^0}}{\braket{\Psi_{\rm \eta}^0}{\Psi_{\rm R}^0}}\ .
    \label{eq:Projector_LR}
\end{align}
where $\eta = \rm L$. Direct application into Eq.~\eqref{Eq:marker} results in the real-space topological marker for the interacting non-Hermitian SSH model [Eq.~\eqref{eq:Ham_chosen}].

In turn, to probe the emergence of the CDW phase, we utilize the finite-size scaling of the charge structure factor, defined as
\begin{align}
    S_{\rm cdw}^{\eta \text{R}}(q) = \frac{1}{N}\sum_{r,r'} \frac{ e^{-i q (r-r')} \langle \Psi_\eta^0 | \hat n_{r} \hat n_{r'} |\Psi_{\rm R}^0 \rangle}{\braket{\Psi_{\rm \eta}^0}{\Psi_{\rm R}^0}}\ ,
    \label{eq:scdw_lr_rr}
\end{align} 
where $\eta ={\rm L,R} $, depending on the choice of the ground state considered to compute the correlations. Here, $r$ and $r'$ denote the position in the lattice, combining the unit cell and the sublattices, and $q$ is the CDW ordering wavevector. However, the choice of eigenvector (left or right) leads to significant differences in the results for charge correlations and local density profiles. In the main text, we always consider $\eta={\rm L}$, but we discuss, for completeness, the results considering $\eta={\rm R}$ in Sec.~\ref{App:left_right_comparison} of the SM \cite{supp}. Nevertheless, we emphasize that both local density and the charge correlations are real-valued quantities for either $\eta=\rm R$ or $\eta=\rm L$.


\begin{figure}[t]
    \centering
    \includegraphics[width=0.47\textwidth]{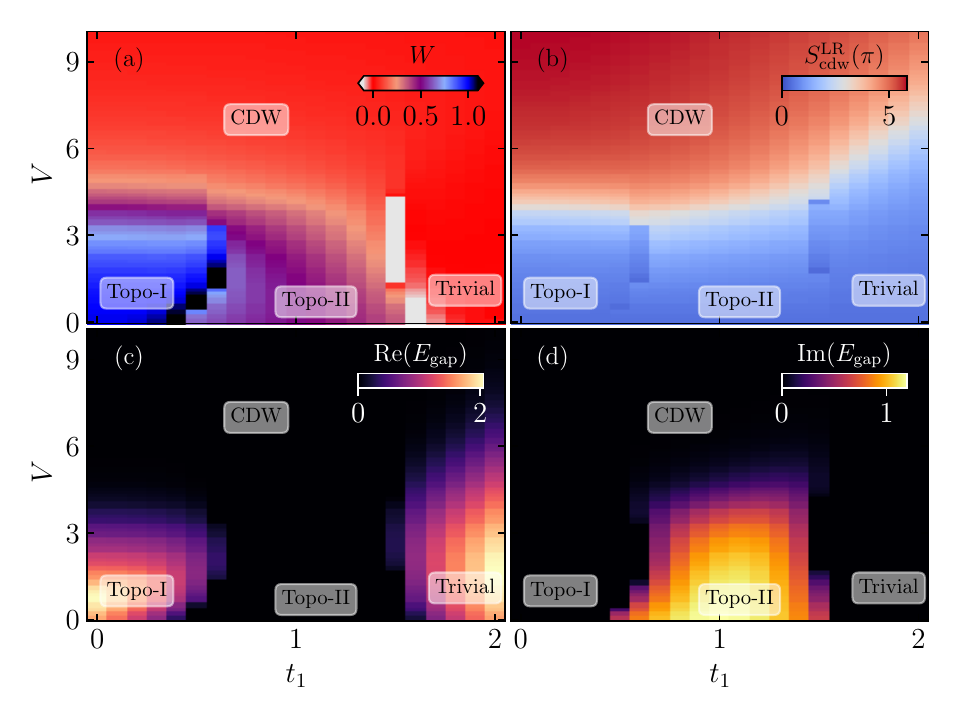}
    \caption{Results in the $V$ \textit{vs} $t_1$ plane, showing (a) the topological marker, (b) the charge structure factor, and (c) the real [(d) imaginary] many-body energy gap. Topo-I and Topo-II stand for the two topological phases with $W=1$ and $W=1/2$, respectively, and the trivial insulator phase possesses $W=0$. Data for $\delta=0.6$, and $N = 24$. Here, $S_{\rm cdw}^{\eta\rm R}(\pi)$ is computed with left and right eigenvectors (i.e., $\eta={\rm L}$).} 
    \label{fig:EDdiagram}
\end{figure}

\begin{table}[t]
\centering
\begin{tabular}{c|cccc}
\hline\hline
 Phase & $\quad\mathrm{Re}(E_{\mathrm{gap}})\quad$ & $\quad\mathrm{Im}(E_{\mathrm{gap}})\quad$ & $\quad m_{\mathrm{cdw}}\quad$  & $\quad W\quad$\\
\hline
Topo-I  & finite & 0  & 0 & 1   \\
Topo-II & 0      & finite   & 0 & 1/2 \\
Trivial   & finite & 0  & 0 & 0   \\
CDW & 0      & 0  & finite & 0   \\
\hline\hline
\end{tabular}
\caption{Characterization of the phases of the system under periodic boundary conditions in terms of the real and imaginary parts of the many-body gap, the CDW structure factor, and the topological marker.}
\label{tab:phase_diagram}
\end{table}
\vskip0.05in
\prlsection{Results} We begin with the PBC results for a representative system size, $N=24$, shown in Fig.~\ref{fig:EDdiagram}. The topological marker $W$, plotted in the plane of intra-cell hopping $t_1$ and nearest-neighbor interaction strength $V$ in Fig.~\ref{fig:EDdiagram}(a), distinguishes three regimes for weak to intermediate interactions, $V\lesssim 3$: a Topo-I regime with $W=1$, a Topo-II regime with $W=1/2$, and a trivial regime with $W=0$. These regimes are continuously connected to those found in the noninteracting limit, as discussed in the SM~\cite{supp}, and their finite-size transition lines remain close to the corresponding $V\to0$ boundaries~\cite{Yin2018,Arouca2020}. To further characterize these regimes, we compute the real and imaginary parts of the many-body gap, ${\rm Re}(E_{\rm gap})$ and ${\rm Im}(E_{\rm gap})$, shown in Figs.~\ref{fig:EDdiagram}(c) and \ref{fig:EDdiagram}(d), respectively. The Topo-I and trivial regimes are associated with a finite real gap, whereas the Topo-II regime is characterized by an imaginary gap. Combining these gap diagnostics with the topological marker yields the finite-size phase boundaries for $V\lesssim 3$, summarized in Fig.~\ref{fig:EDdiagram}(b).

As the interaction strength $V$ increases, charge correlations are enhanced and eventually drive the system into a CDW phase, as shown in Fig.~\ref{fig:EDdiagram}(b) through the staggered, $q=\pi$, CDW structure factor $S^{\rm LR}_{\rm cdw}(\pi)$. The onset of charge order preempts the topological regimes: in the CDW phase, the topological marker vanishes, and neither the real nor the imaginary many-body gap provides a finite topological signature [see Figs.~\ref{fig:EDdiagram}(a), (c), and (d)]~\cite{Melo2023}. Although the nearest-neighbor interaction itself does not explicitly break chiral symmetry, the CDW phase spontaneously breaks it, leading to a finite order parameter in the thermodynamic limit. Consequently, the chiral-symmetry-protected topological marker becomes trivial in the charge-ordered phase. This competition between charge order and chiral-symmetry-protected topology can be understood intuitively from the mean-field analysis discussed in Sec.~\ref{App:MFA} of the SM~\cite{supp}. However, the mean-field treatment does not quantitatively reproduce the full phase diagram. In particular, it fails to capture the suppression of the Topo-II phase by interactions and the resulting interaction-driven transitions between the Topo-II and Topo-I phases~\cite{supp}. The resulting PBC phase diagram is summarized in Table~\ref{tab:phase_diagram}.

\begin{figure}[t]
    \centering
    \includegraphics[scale = 0.5]{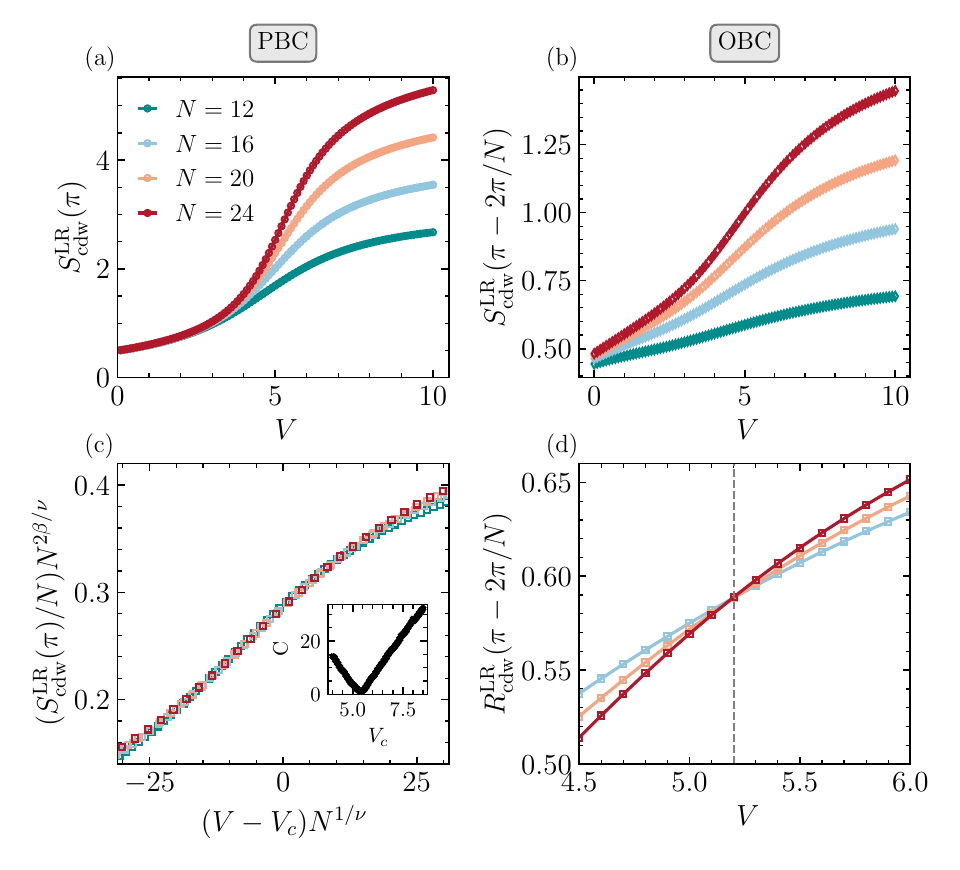}
    \caption{The CDW structure factor at (a) $q = \pi$ under PBC and (b) at $q = \pi - 2\pi/N$ under OBC. (c) Finite-size-collapsed CDW structure factor under PBC; the critical interaction $V_c = 5.4 \pm 0.1$ is obtained by minimizing the collapse cost function (inset). (d) CDW correlation ratio (see text) as a function of $V$ under OBC, where the dashed line marks the crossing point $V_c \approx 5.2$ for different system sizes. Data are for $t_1 = 1.6$, and $\delta = 0.6$, obtained using left and right eigenvectors.}
    \label{fig:lr_data_collapse}
\end{figure}

To locate the CDW transition more accurately, we perform a finite-size scaling analysis of $S^{\rm LR}_{\rm cdw}(\pi)$ as a function of $V$. Since the CDW order parameter, $m_{\rm cdw}\equiv|\langle \hat n_A-\hat n_B\rangle|$, breaks a discrete $\mathbb{Z}_2$ symmetry, we assume the $(1+1)$D Ising universality class, with critical exponents $\beta=1/8$ and $\nu=1$~\cite{Cardy1996, Sachdev2011}. In the thermodynamic limit, $S_{\rm cdw}/N$ approaches $m_{\rm cdw}^2$, motivating the scaling ansatz~\cite{Melo2023}
\begin{align}
    S_{\rm cdw}^{\rm LR}(\pi)/N = N^{-2\beta/\nu} f[(V - V_{c})N^{1/\nu}]\ ,
    \label{eq:scaling_ansatz}
\end{align}
in the critical region. We determine $V_c$ by minimizing a collapse cost function, $C = \sum_j (|y_{j+1} - y_j|)/(\max\{y_j\} - \min\{y_j\})-1$~\cite{Suntajs2020, Jin2022, Mondaini2023}, where $y_j$ are the rescaled values of $S_{\rm cdw}/N$, ordered by the scaling variable $(V-V_c)N^{1/\nu}$~\cite{Suntajs2020,Jin2022,Mondaini2023}. The resulting collapse is shown in Fig.~\ref{fig:lr_data_collapse}(c), with the minimization of $C$ displayed in the inset, yielding $V_c=5.4\pm0.1$.

\begin{figure}[t]
    \centering
    \includegraphics[scale = 0.5]{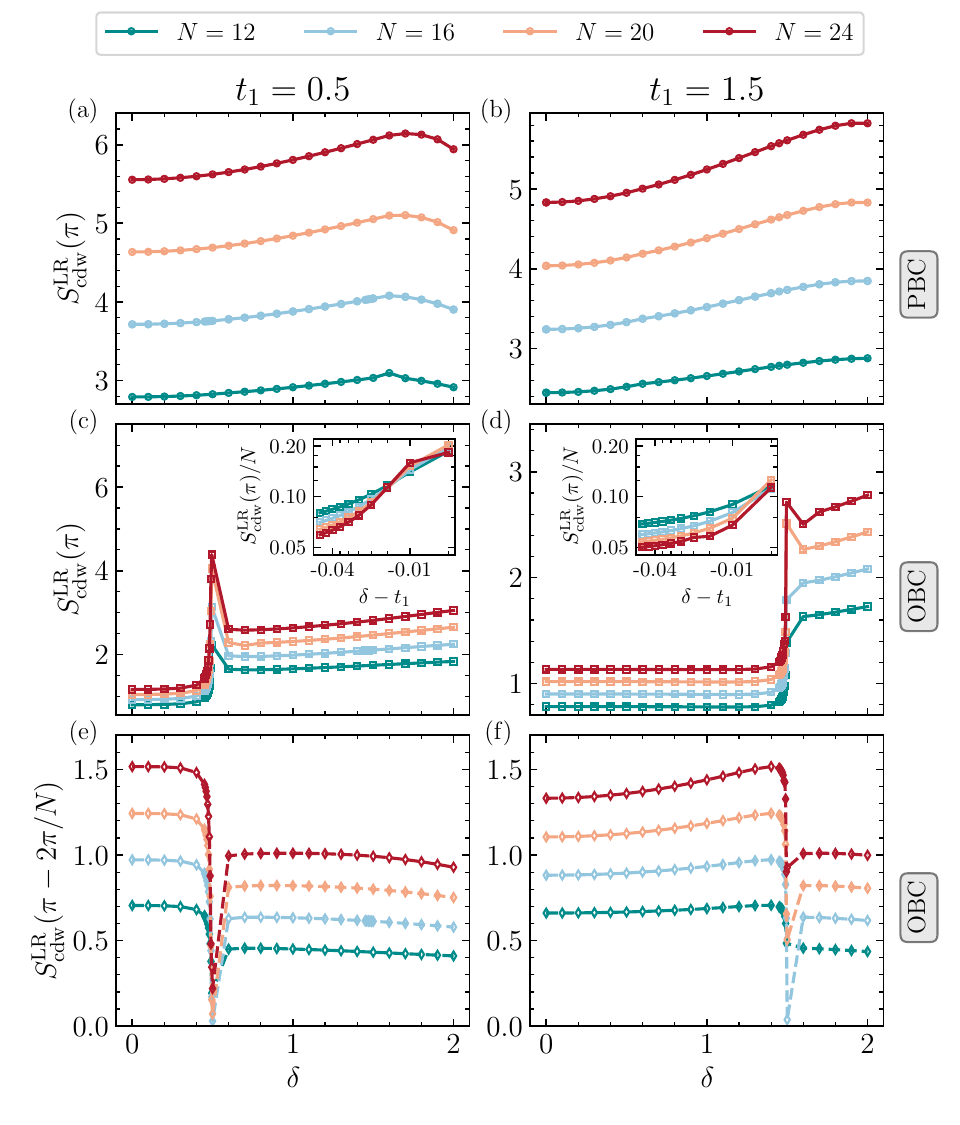}
    \caption{Charge structure factor as a function of $\delta$ for several lattice sizes for $V = 8$, and $t_1 = 0.5$ (left panels), and $t_1 = 1.5$ (right panels); (a) and (b) present data for PBC, while (c) and (d) display data for OBC, all for $q=\pi$. The insets in panels (c) highlight the enhanced region. Panels (e) and (f) display results for OBC this time for $q=\pi-2\pi/N$.}
    \label{fig:cdw_cuts_delta}
\end{figure}

For OBC, translational symmetry is explicitly broken, and the two symmetry-related, staggered CDW configurations leading to vanishing gaps are no longer degenerate in finite systems, resulting in a unique, boundary-pinned ground state. Although the CDW phase still corresponds to a discrete $\mathbb{Z}_2$ ordering in the thermodynamic limit, identifying it in finite systems via a data collapse using the Ising-model critical exponents is more challenging. Instead, we determine the CDW phase boundaries through the correlation ratio~\cite{Kaul2015},
\begin{align}
    R_{\rm cdw}^{\eta\rm R}(q) = 1- \frac{S_{\rm cdw}^{\eta\rm R}(q + \delta q)}{S_{\rm cdw}^{\eta\rm R}(q)},
    \label{eq:RgammaLV}
\end{align}
where, $q  = \pi - 2\pi/N $ and $\delta q = 2\pi/N$. For a CDW phase, the charge correlations are peaked at the ordering vector, yielding $R_{\rm cdw}(q) \to  1$ as the system size increases. In the other phases, the peak is suppressed for larger system sizes and produces $R_{\rm cdw}(q) \to 0$. Therefore, at the critical point, the correlation ratio is independent of $N$, so that plots of $R_{\rm cdw}(q)$ for different $N$ cross, providing estimates for the critical $V_c$. Figures~\ref{fig:lr_data_collapse}(b) and \ref{fig:lr_data_collapse}(d) display the charge structure factor and the correlation ratio as a function of $V$ for several system sizes under OBC. In the region where $S_{\rm cdw}^{\rm LR}$ presents an inflection point [Fig.~\ref{fig:lr_data_collapse}(b)], the correlation ratio curves cross at $V\sim 5.2$ [Fig.~\ref{fig:lr_data_collapse}(d)]. After collecting the critical values of $V$ for PBC and OBC from the data collapse and correlation ratio crossing, respectively, we determine the CDW transition lines presented in Figs.~\ref{fig:lattice}(b) and \ref{fig:lattice}(c). As noted before, due to the non-Hermitian skin effect, non-Hermitian Hamiltonians in general have different spectra and, consequently, different phase diagrams for PBC and OBC \cite{Yao2018SSH}.

In the Hermitian limit, increasing the dimerization by taking $t_1\neq t_2$ opens a single-particle gap and suppresses the charge susceptibility associated with the Peierls instability at $q=2k_{\rm F}$~\cite{Peierls1955,Zhu2015,Zhu2017}. Consequently, the critical interaction required to stabilize CDW order increases. In the non-Hermitian case, however, the nonreciprocity controlled by $\delta$ modifies the noninteracting spectrum and can bring the system close to exceptional points~\cite{Arouca2020}. In this regime, the spectral response to perturbations is enhanced, making interaction effects more pronounced~\cite{Arouca2023}. We therefore examine how the CDW correlations evolve with $\delta$ and show that nonreciprocity can enhance the tendency toward charge ordering.

The evolution of $S_{\rm cdw}^{\rm LR}(q)$ with nonreciprocity is shown in Fig.~\ref{fig:cdw_cuts_delta} for both PBC and OBC. Under PBC, the charge correlations exhibit only a modest enhancement over a broad range of $\delta$ [Figs.~\ref{fig:cdw_cuts_delta}(a) and \ref{fig:cdw_cuts_delta}(b)]. By contrast, under OBC the staggered structure factor increases sharply as the system approaches the exceptional point, $\delta=t_1$ [Figs.~\ref{fig:cdw_cuts_delta}(c) and \ref{fig:cdw_cuts_delta}(d)]. The insets show the normalized structure factor in a magnified window near the exceptional point on a log-log scale, emphasizing the enhanced charge response. This enhancement is most pronounced in the $q=\pi$ component; the nearby finite-size momentum $q=\pi-2\pi/N$ instead decreases near the exceptional point under OBC [Figs.~\ref{fig:cdw_cuts_delta}(e) and \ref{fig:cdw_cuts_delta}(f)].

It is important to emphasize that finite-size ED becomes numerically delicate near the exceptional line, $\delta=t_1$, where the low-lying spectrum collapses, and both the real and imaginary parts of the eigenenergies become strongly clustered as $\delta\to t_1$ (see Sec.~\ref{App:pt_symm_eigen} of the SM~\cite{supp}). Within this limitation, our OBC results indicate that the exceptional line separates two charge-ordered regimes, a staggered CDW phase dominated by the $q=\pi$ component and a CDW phase whose dominant correlations occur at a nearby finite-size wave vector, $q<\pi$. We also stress that, although increasing $\delta$ enhances charge correlations near the exceptional line, it does not substantially lower the critical interaction strength required to stabilize CDW order. This point is further supported by the comparison between left-right and right-right correlation functions discussed in Sec.~\ref{App:left_right_comparison} of the SM~\cite{supp}.

\begin{figure}[t]
    \centering
    \includegraphics[scale = 0.5]{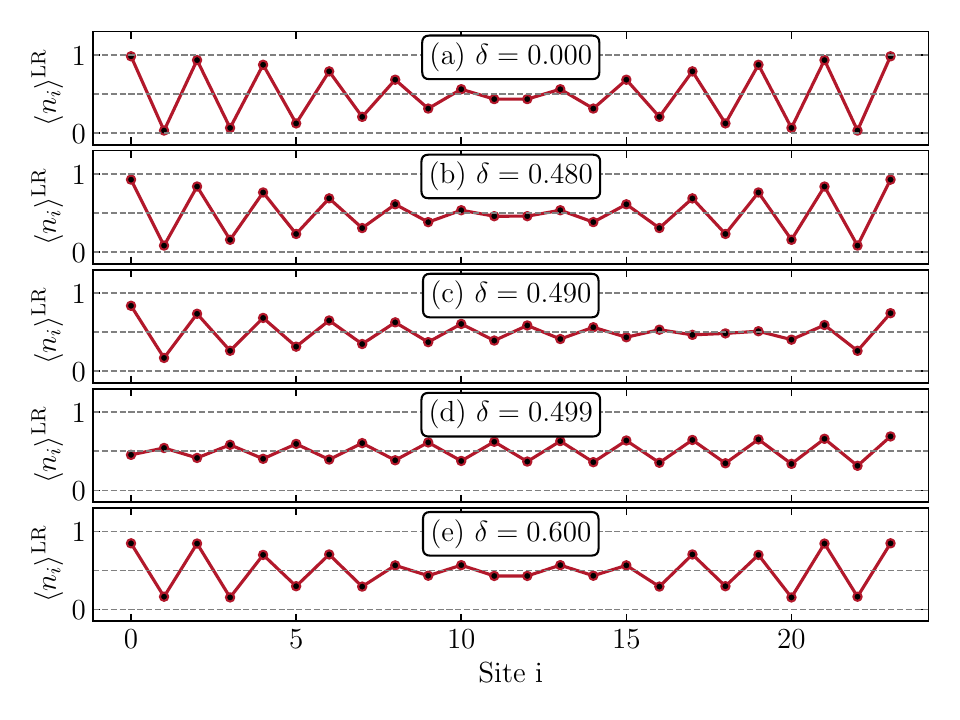}
    \caption{Local density under OBC for $L=24$, $V=8$, $t_1=0.5$, for different values of $\delta$. All data computed using the left-right (LR) convention.}
    \label{fig:local_density_obc_lr}
\end{figure}

To further assess the effect of the exceptional point on charge correlations, we examine the local density profile under OBC inside the CDW regime. Figure~\ref{fig:local_density_obc_lr} shows the biorthogonal density $\langle n_i\rangle^{\rm LR}$ for several values of $\delta$. Because OBC explicitly breaks translational invariance, the density profile is spatially inhomogeneous even away from the exceptional point. In particular, the central sites are less occupied than the edges. As $\delta\to t_1$, the density modulation becomes more strongly staggered, consistent with the enhancement of the $q=\pi$ structure factor shown in Figs.~\ref{fig:cdw_cuts_delta}(c) and \ref{fig:cdw_cuts_delta}(d). However, the LR density remains approximately mirror-symmetric about the center of the chain and does not exhibit a robust kink-like domain wall.

This behavior highlights an important distinction between biorthogonal LR observables and right-right RR observables in non-Hermitian systems. If the density is computed using only right eigenvectors, as in Ref.~\cite{Zhong2025}, OBC can produce a strongly asymmetric kink-like texture. In our LR observables, this asymmetry is absent, and the density profile recovers the expected mirror symmetry of the finite chain. Moreover, RR expectation values can generate charge modulations even under PBC, where such spatial structure should not arise from boundary pinning. These differences indicate that the kink-like pattern is tied to the nonorthogonality and skin accumulation of right eigenstates, rather than to a distinct biorthogonal CDW-kink phase. A detailed comparison between LR and RR observables is provided in Sec.~\ref{App:left_right_comparison} of the SM~\cite{supp}.


\prlsection{Discussions and conclusions} We investigated the effect of interactions on the topological phases of the non-Hermitian SSH model using exact diagonalization. By combining a real-space topological marker with charge correlations and the complex many-body spectrum, we showed that charge ordering competes with and ultimately destroys the chiral-symmetry-protected topological regimes. In particular, interactions modify the competition between the Topo-I phase, characterized by $W=1$, and the intrinsically non-Hermitian Topo-II phase, characterized by $W=1/2$. This interaction-driven redistribution of the topological regimes is not captured by the mean-field treatment discussed in the SM~\cite{supp}. We also characterized the onset of CDW order and found scaling behavior consistent with the $(1+1)$D Ising universality class, as in the Hermitian interacting SSH chain~\cite{Melo2023}. Finally, we showed that nonreciprocal hopping can enhance charge correlations and reshape the local density profile, particularly under OBC, where the response is strongly affected by the proximity to the exceptional line of the corresponding noninteracting problem.

This work opens interesting perspectives for non-Hermitian physics. One is to consider the computation of this invariant, considering a full dissipative setting, using the steady state instead of the many-body ground state. Another interesting direction is to investigate phases protected by (non-Hermitian) time-reversal and particle-hole symmetries using markers together with measures such as entanglement spectrum and entropy, which can also track ``fractionalized'' phases~\cite{Li2008, Zaletel2013}.

\textit{Note added.} At the final stages of completing this work, we became aware of Ref.~\cite{Lu2026}, which studies the same interacting non-Hermitian SSH Hamiltonian using a generalized-Brillouin-zone construction. Where directly comparable, our results are consistent regarding the emergence of CDW order. However, Ref.~\cite{Lu2026} identifies a topological-CDW regime at strong interactions by combining the CDW structure factor with a many-body biorthogonal Zak phase. This differs from our characterization based on the real-space topological marker, which becomes trivial once CDW order develops. We attribute this difference to the choice of topological diagnostic: as shown already in the Hermitian interacting SSH chain, the many-body Zak phase can remain quantized across a transition into a charge-ordered insulating phase and therefore may not distinguish the topological-to-CDW transition captured by the real-space marker~\cite{Melo2023}.

\section*{Acknowledgments} S.A.S.-J. gratefully acknowledges the Brazilian agency CNPq for funding support, grant No.~201000/2024-5. PBM acknowledges the support of the Brazilian agencies CAPES, Finance code 001, and the Brazilian agency CNPq for funding support, grant No. 140264/2026-4. R.M.~acknowledges support from the T$_{\rm c}$SUH Welch Professorship Award. Part of the calculations used resources from the Research Computing Data Core at the University of Houston. This work also used TAMU ACES at Texas A\&M HPRC through allocation PHY240046 from the Advanced Cyberinfrastructure Coordination Ecosystem: Services \& Support (ACCESS) program, which is supported by U.S. National Science Foundation grants 2138259, 2138286, 2138307, 2137603, and 2138296. RA acknowledges the support of the INCT project Advanced Quantum Materials, involving the Brazilian agencies CNPq (Proc.
408766/2024-7), FAPESP (Proc. 2025/27091-3), and CAPES

\bibliographystyle{apsrev4-2mod}
\bibliography{References}



\clearpage
\newpage

\begin{onecolumngrid}
\phantomsection
\label{sec:sm}
\begin{center}
{\fontsize{12}{12}\selectfont
		\textbf{Supplemental Material for ``Enhancement of charge correlations and real-space topological marker on an interacting non-Hermitian Su-Schrieffer-Heeger model''\\[5mm]}}
	{\normalsize Sebasti\~ao dos A. Sousa-Júnior\,\orcidlink{0000-0002-4266-3780}$^{1}$, Pedro B. Melo\,\orcidlink{0000-0002-3726-761X}$^{2,3}$, Rubem Mondaini\,\orcidlink{0000-0001-8005-2297}$^{1,4}$, Arnob Kumar Ghosh\,\orcidlink{0000-0003-0990-8341}$^{5}$, Rodrigo~Arouca\,\orcidlink{0000-0003-4214-1437}$^{6}$ \\[1mm]}
	{\small \textit{$^1$Department of Physics, University of Houston, Houston, Texas 77204, USA}\\[0.5mm]}
	{\small \textit{$^2$Departamento de F\'isica, PUC-Rio, 22452-970, Rio de Janeiro RJ, Brazil}\\[0.5mm]}
    {\small \textit{$^3$Universit\`a degli Studi di Palermo, Dipartimento di Fisica e Chimica - Emilio Segr\`e, via Archirafi 36, I-90123 Palermo, Italy}\\[0.5mm]}
    {\small \textit{$^4$Texas Center for Superconductivity, University of Houston, Houston, Texas 77204, USA}\\[0.5mm]}
    {\small \textit{$^5$Department of Physics and Astronomy, Uppsala University, Box 524, 75120 Uppsala, Sweden}\\[0.5mm]}
    {\small \textit{$^6$Brazilian Center For Research in Physics, Rua Doutor Xavier Sigaud 150, Rio de Janeiro, 22290-180, Brazil}\\[0.5mm]}
\end{center}

\vspace{-0.1cm}
\normalsize
\begin{center}
	\parbox{14cm}{In this Supplemental Material, we first discuss the effective non-Hermitian Hamiltonian derived from the Lindbladian description in Sec.~\ref{app:Lind}. Then, in Sec.~\ref{subsec:NIS}, we provide the topological marker and phase diagram for the non-interacting regime. We also present a mean-field treatment of the interacting model in Sec.~\ref{App:MFA}, additional results on $\mathcal{P}\mathcal{T}$-symmetry breaking in Sec.~\ref{App:pt_symm_eigen}, and the phase diagram under open boundary conditions in Sec.~\ref{App:OBC_phase_diag}. Furthermore, we also compare results with a different eigenstate convention used to compute physical observables in Sec.~\ref{App:left_right_comparison}. Finally, we conclude with a discussion on topological phase transition under weak interaction in Sec.~\ref{app:weak_coupling}.}
\end{center}

\setcounter{page}{1}

\setcounter{section}{0}
\setcounter{equation}{0}
\setcounter{figure}{0}
\setcounter{table}{0}

\setcounter{section}{0}
\setcounter{equation}{0}
\setcounter{figure}{0}
\setcounter{table}{0}

\renewcommand{\thesection}{S\arabic{section}}
\renewcommand{\theequation}{S\arabic{equation}}
\renewcommand{\thefigure}{S\arabic{figure}}
\renewcommand{\thetable}{S\arabic{table}}

\end{onecolumngrid}

\begin{twocolumngrid}

\section{Lindbladian description of the non-Hermitian effective Hamiltonian}
\label{app:Lind}

To motivate the adopted non-Hermitian effective Hamiltonian in Eq.~(1) in the main text, we provide its description from a quantum open-systems perspective. We consider the post-selection limit, in which the system's non-unitary time evolution occurs between the quantum jumps. The Hamiltonian $\hat {\mathcal H}$ emerges for the evolution of conditioned no-jump trajectories. We consider a system coupled to an external environment, where the evolution of the density matrix $\rho(t)$ is governed by the Lindblad master equation~\cite{BreuerPetruccione2002}
\begin{equation}
    \frac{{\rm d}\hat \rho(t)}{{\rm d}t} = {\cal L}\left[\hat \rho(t)\right].
\end{equation}
The Liouvillian superoperator $\mathcal{L}$ is defined as
\begin{equation}
    \mathcal{L}[\hat{\rho}] = -i[\hat{\mathcal H_0}, \hat{\rho}]
    +\sum_j \mathcal{D}[\hat{L}_j](\hat{\rho}),
\end{equation}
where $\hat{\mathcal H_0}$ is the Hermitian Hamiltonian of the system and
$\{\hat{L}_j\}$ are the jump operators describing the microscopic coupling
to the environment. In our case, the Hamiltonian $\hat{\mathcal H_0}$ represents an interacting Su-Schrieffer-Heeger~(SSH) model. The dissipator $\mathcal{D}[\hat{L}_j]$ acts on the
density operator as
\begin{equation}
    \mathcal{D}[\hat{L}_j](\hat{\rho}) =
    \hat{L}_j \hat{\rho} \hat{L}_j^{\dagger}
    -\frac{1}{2}\{\hat{L}_j^{\dagger}\hat{L}_j, \hat{\rho}\}.
\end{equation}
The effective non-Hermitian Hamiltonian, $\hat {\mathcal H}$, governs the time evolution of the system between quantum jumps. It is obtained by combining the coherent evolution with the anti-Hermitian decay term:
\begin{equation}
    \hat {\mathcal H} = \hat{\mathcal H}_0 - \frac{i}{2} \sum_j \hat L_j^\dagger \hat L_j^{\phantom{\dagger}} \ .
    \label{eq:NonH-Ham_eff}
\end{equation}
To realize the specific non-reciprocity of the non-Hermitian SSH model, considered in this work, we introduce dissipation acting solely on the intracell degrees of freedom. We define the jump operators as
\begin{equation}
    \hat L_j = \sqrt{2\delta} (\hat c_{j,A} + i\hat c_{j,B})\ . \label{eq:GKSL_operators}
\end{equation}
The Hermitian part of the dynamics is given by the interacting SSH Hamiltonian:
\begin{align}
    \hat {\mathcal H}_0 = \sum_{j} \Big[ &t_1 \hat c^\dagger_{j,B} \hat c_{j,A}^{\phantom{\dagger}} + t_2 \hat c^\dagger_{j+1,A}\hat c_{j,B}^{\phantom{\dagger}} + \text{H.c.} \nonumber \\
    &+ V( \hat n_{j,A} \hat n_{j,B} + \hat n_{j,B} \hat n_{j+1,A}) \Big]\ .
\end{align}
Substituting Eq.~(\ref{eq:GKSL_operators}) into the decay term yields
\begin{align}
    \hat L_j^\dagger \hat L_j^{\phantom{\dagger}} &= 2\delta (\hat c_{j,A}^{\dagger} - i\hat c_{j,B}^{\dagger})(\hat c_{j,A}^{\phantom{\dagger}} + i\hat c_{j,B}^{\phantom{\dagger}}) \nonumber \\
    &=2\delta [\hat n_{j,A} + \hat n_{j,B} + i(\hat c_{j,A}^{\dagger}\hat c_{j,B}^{\phantom{\dagger}} - \hat c_{j,B}^{\dagger}\hat c_{j,A}^{\phantom{\dagger}})].
\end{align}
Finally, the effective non-Hermitian in Eq.~(\ref{eq:NonH-Ham_eff}) takes the form
\begin{align}
    \hat {\mathcal H} = \sum_{j} \Big[ &t_1 \hat c^\dagger_{j,B} \hat c_{j,A}^{\phantom{\dagger}} + t_2 \hat c^\dagger_{j+1,A}\hat c_{j,B}^{\phantom{\dagger}} + \text{H.c.} \nonumber \\
    &+ V( \hat n_{j,A} \hat n_{j,B} + \hat n_{j,B} \hat n_{j+1,A}) \nonumber \\
    &- i\delta (\hat n_{j,A} + \hat n_{j,B}) + \delta (\hat c^\dagger_{j,A} \hat c_{j,B}^{\phantom{\dagger}} - \hat c^\dagger_{j,B}\hat c_{j,A}^{\phantom{\dagger}}) \Big].
    \label{eq:eq_Eff}
\end{align}
The dissipation introduces an imaginary onsite potential $-i\delta$, which causes a global decay of the norm, and a non-reciprocal hopping term that breaks Hermiticity. Combining the non-reciprocal hopping term with the Hermitian intracell hopping produces asymmetric hopping amplitudes, $t_1+\delta$ and $t_1-\delta$, thereby realizing non-reciprocity studied in the main text. In a fixed particle-number sector, the onsite term $-i\delta\sum_j(\hat n_{j,A}+\hat n_{j,B})$ is proportional to the total particle number and therefore only produces an overall decay of the no-jump wave function. It can consequently be removed when considering normalized states or spectral properties that are insensitive to a uniform imaginary energy shift, in particular for those in equilibrium, our main focus.

\section{Topological marker for the non-interacting system}
\label{subsec:NIS}

To probe the topology of non-Hermitian topological invariant in one dimension for single-particle sectors of the Hilbert space, one uses the non-Hermitian winding number~\cite{Yin2018}. Considering periodic boundary conditions (PBC), the Bloch Hamiltonian of a two-band model with chiral symmetry is given by $h(k) = \vec{d}(k)\cdot\vec{\sigma}$, where the vector $\vec{d}(k) = (d_x,d_y)$ is multiplied by the Pauli matrix vector $\vec{\sigma} = (\sigma_x,\sigma_y)$. The non-Hermitian (NH) topological winding number is given by $\textbf{W} = (W_1,W_2)$, where $W_i$ ($i = 1,2$) is given by~\cite{Yin2018, Arouca2020}
\begin{equation}
    W_i = \frac{1}{2\pi}\oint_C \partial_k\arctan \left[\frac{\text{Re}(d_y(k)) \pm \text{Im}(d_x(k))}{\text{Re}(d_x(k)) \mp \text{Im}(d_y(k))}\right] {\rm d}k\ ,
\label{eq:winding_number}    
\end{equation}
with the upper (lower) sign for $i = 1$ ($i = 2$), and where $\text{Re}(d_i(k))$ and $\text{Im}(d_i(k))$ refers to the real and imaginary parts of $d_i(k)$, respectively. The winding number is given by $W = W_1 + W_2$, which in Hermitian phases is either $W = 0$ (trivial phase) or $W = 1$ (topological phase). In the non-Hermitian topological phase it is given by $W = 1/2$, being either $\textbf{W} = (1/2, 0)$ or $\textbf{W} = (0, 1/2)$. Notably, due to the presence of the non-Hermitian skin effect, this invariant is not directly related to the presence of edge states, and one needs to use either the biorthogonal polarization \cite{Kunst18} or the generalized Brillouin zone formalism \cite{Yao2018SSH} to characterize the topology of the system with OBCs.

\begin{figure}[t]
    \centering
    \includegraphics[scale = 0.5]{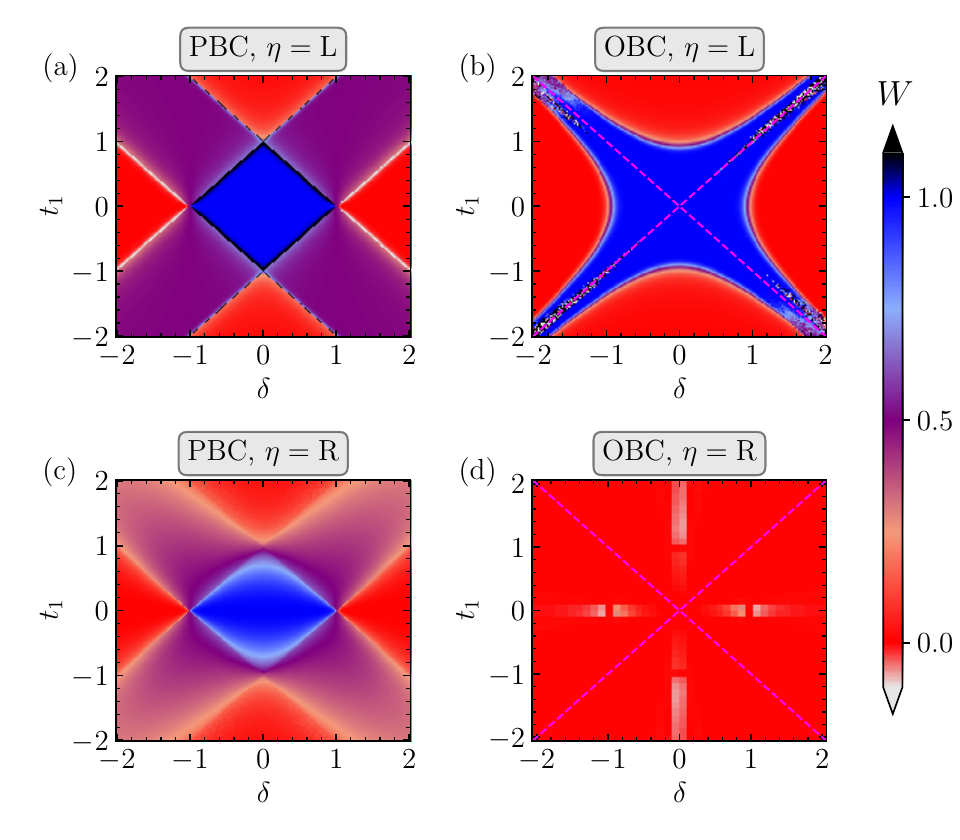}
    \caption{Topological marker $W$ as a function of $t_1$ and $\delta$ for $N = 60$, $t_2 = 1$, and $V = 0$ under PBC (a) and OBC (b), obtained from left and right eigenvectors, i.e., with $\eta = \rm L$. Panels (c) and (d) show the corresponding results calculated using only right eigenvectors, $\eta = \rm R$. The dashed lines in (b) and (d) mark the boundary between Hermitian and non-Hermitian topological regions under OBCs~\cite{Arouca2020}.}
    \label{fig:non_interacting_diagrams}
\end{figure}

The non-interacting case of the NHSSH \cite{Yao2018SSH, Yin2018, Arouca2020} model presents three distinct phases for PBC, a trivial phase where the winding number vanishes, a `Topo-I' phase where the winding number is $W=1$, and a phase where the winding number is $W=1/2$, dubbed `Topo-II'. As a benchmark for our simulations, we tested the topological marker defined in Ref.~\cite{Song2019} and in the main text under both periodic and open boundary conditions for $N=60$ sites, as is shown in Fig.~\ref{fig:non_interacting_diagrams}. Notably, the marker is computed via the matrix elements of the projector operator:
\begin{align}
    P_{rr^\prime} = \sum_{\nu\in S} \frac{\bra{\psi_{\eta}^\nu} \hat c_r^\dagger \hat c_{r^\prime}^{\phantom{\dagger}} \ket{\psi_{\rm R}^\nu}}{\braket{\psi_{\eta}^\nu}{\psi_{\rm R}^\nu}}\ ,
    \label{eq:projector}
\end{align}
where $\eta = \text{L}$ in Eq.~\eqref{eq:projector} of the main text. Indeed, for $\eta = \rm L$, and under PBCs, the topological marker reproduces the phase diagram in the $t_1 \mhyphen \delta$ plane with great accuracy. Under OBC, on the other hand, numerical instabilities are more pronounced near the higher-order exceptional points, leading to noise in the topological marker value, yet the phase diagram resembles that of Ref.~\cite{Arouca2020}. 

Now, if one instead uses a definition of the projector with $\eta = \rm R$ for the evaluation of the topological marker, one fails to reproduce the correct phase diagrams. This is illustrated in Figs.~\ref{fig:non_interacting_diagrams}(c) and \ref{fig:non_interacting_diagrams}(d). Under PBCs, quantization is suppressed, although the resulting phase diagram still qualitatively reproduces the phase boundaries. Under OBC, however, the topological marker is strongly degraded, and phase boundaries cannot be resolved in the $t_1 \mhyphen \delta$ parameter space.

It is worth noting that in the two-dimensional systems, the topological invariant (Chern number) for non-Hermitian systems is the same for both $\eta = \rm L$ and $\eta = \rm R$~\cite{shen2018}. Here, however, our one-dimensional model yields the correct topological marker only for $\eta = L$.

\subsection{Equivalence of topological markers}

It is important to emphasize that the topological marker used in the main text, Eq.~\eqref{Eq:marker}, is equivalent to the open-bulk winding marker introduced in Ref.~\cite{Song2019}. The latter is written as
\begin{equation}
    W^{\prime} = \frac{1}{2N_{\rm C}} \mathrm{Tr}_{\rm C}
    \bigl[\hat S \hat Q^{\prime} [\hat Q^{\prime},\hat X] \bigr]\ ,
    \label{eq:song_marker}
\end{equation}
where $\mathrm{Tr}_{\rm C}$ denotes a trace restricted to the central region of the chain. The only difference is notational. In our convention, $\hat P$ projects onto one member of each chiral pair and $\hat Q=\mathbb{1}-\hat P$ is the complementary projector. In Ref.~\cite{Song2019}, the corresponding object entering the marker is instead the flattened chiral operator
\begin{equation}
    \hat Q^{\prime} =\hat P-\hat Q = 2\hat P-\mathbb{1}\ .
\end{equation}

To show the equivalence, first note that
\begin{equation}
    [\hat Q^{\prime},\hat X]
    =[\hat P,\hat X]-[\hat Q,\hat X] \
    = 2[\hat P,\hat X]\ .
\end{equation}
Therefore,
\begin{align}
    \hat Q^{\prime}[\hat Q^{\prime},\hat X]
    &=
    2(\hat P-\hat Q)
    (\hat P\hat X-\hat X\hat P) \notag \\
    &=
    2\bigl(
    \hat P\hat X
    -\hat P\hat X\hat P
    +\hat Q\hat X\hat P
    \bigr)\ .
\end{align}

Using $\hat P^2=\hat P$, $\hat Q^2=\hat Q$, and
$\hat P\hat Q=\hat Q\hat P=0$, the first two terms become
\begin{equation}
    \hat P\hat X-\hat P\hat X\hat P =
    \hat P\hat X(\mathbb{1}-\hat P) =
    \hat P\hat X\hat Q\  .
\end{equation}
Thus,
\begin{equation}
    \hat Q^{\prime}[\hat Q^{\prime},\hat X] =
    2\bigl( \hat P\hat X\hat Q
    + \hat Q\hat X\hat P
    \bigr)\ .
\end{equation}
Substituting this result into Eq.~\eqref{eq:song_marker} gives
\begin{align}
    W^{\prime} = \frac{1}{N_{\rm C}} \mathrm{Tr}_{\rm C}
    \Bigl[\hat S\hat P\hat X\hat Q \ +
    \hat S\hat Q\hat X\hat P \Bigr], 
\end{align}
which is precisely the marker used in the main text. Therefore, the two forms are algebraically equivalent. The only remaining freedom is the convention used to choose which member of each chiral pair is included in $\hat P$; reversing this choice changes the overall sign of the winding number.

\section{Mean Field Approximation}
\label{App:MFA}

Here, we present a mean-field approach when one considers a nearest-neighbor density-density interaction of the form
\begin{equation}
\hat H_V = V \sum_{\langle r,r' \rangle} \hat n_r \hat n_{r'},
\label{eq:HV_full}
\end{equation}
where the indices $r$ and $r'$ denote different sites in the lattice, i.e., they combine information about the sublattice and the unit cell.
To proceed, we perform a standard mean-field (Hartree) decoupling, neglecting quadratic fluctuations:
\begin{equation}
\hat n_r \hat n_{r'} 
\simeq 
\hat n_r  \langle \hat n_{r'} \rangle
+ \langle \hat n_{r}  \rangle \hat n_{r'} 
- \langle \hat n_{r'}  \rangle \langle \hat n_{r'} \rangle\ .
\label{eq:meanfield_decoupling}
\end{equation}
In a bipartite lattice, we denote the sublattice densities as
\begin{equation}
\langle \hat n_A \rangle = n + \delta_n, 
\qquad
\langle \hat n_B \rangle = n - \delta_n,
\end{equation}
where $\delta_n$ quantifies the charge imbalance between the two sublattices, and $n = 1/2$ at half-filling.
Substituting Eq.~\eqref{eq:meanfield_decoupling} into Eq.~\eqref{eq:HV_full}, each site experiences a local potential proportional to the mean occupation of its nearest neighbors. In the sublattice spinor basis 
$\Psi_k^\dagger = (\hat c_{k,A}^\dagger, \hat c_{k,B}^\dagger)$, the Hamiltonian reads,
\begin{equation}
\hat{\cal H}_{\rm MF} = \sum_k \left(\hat c_{k,A}^\dagger, \hat c_{k,B}^\dagger\right) \hat{h} (k) 
\begin{pmatrix}
\hat c_{k,A} \\
\hat c_{k,B} 
\end{pmatrix} + NV\delta_n^2 ,
\qquad \\
\end{equation}
where,
\begin{equation}
\hat{h}(k) =
\begin{pmatrix}
-M & f^+(k) \\
f^-(k) & M
\end{pmatrix},
\qquad 
M = -\,2V\,\delta_n,
\label{eq:hamiltonian_mf}
\end{equation}
and $f^{\pm}(k) = t_2  e^{\pm i  k} + t_1 \pm \delta$. This matrix produces eigenvalues,
\begin{equation}
E_{\pm}(k) = \pm \sqrt{M^2 + f^+(k) f^-(k)},
\end{equation}
which can be complex. The charge imbalance in the \textit{left-right} conventions reads,
\begin{figure}[t]
    \centering
    \includegraphics[scale = 0.5]{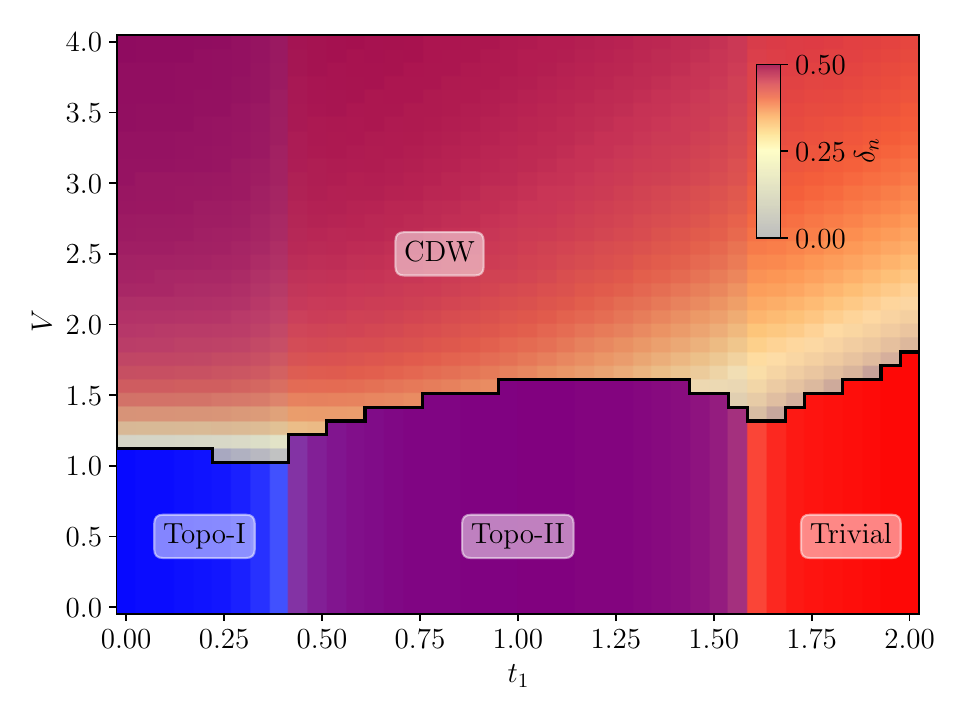}
    \caption{Mean field phase diagram. Data with fixed hopping parameters $\delta=0.6$, $t_2 = 1$, as in Fig.~\ref{fig:EDdiagram} of the main text. Within the CDW phase, the color maps the order-parameter value.}
    \label{fig:MFdiagram}
\end{figure}
\begin{equation}
    \delta_n = \frac{1}{2N}\sum_k \frac{\bra{\psilk} \hat S\ket{\psirk}}{\bra{\psilk} \ket{\psirk}}  
\end{equation}
which results in a self-consistent relation for $\delta_n$, namely
\begin{equation}
   F(\dn) =\dn- \frac{1}{N} \sum_k \frac{V\dn}{\sqrt{4V^2\dn^2 + f^+(k)f^-(k)}} =0.
   \label{Eq:self_consist_MF}
\end{equation}
At zero temperature, the mean field solution then consists of solving Eq.~\eqref{Eq:self_consist_MF} while minimizing the total energy per site,
\begin{equation}
    E = \frac{1}{N} \sum_k E_-(k)  + V\dn^2.
    \label{Eq:mf_energy}
\end{equation}
Figure~\ref{fig:MFdiagram} shows the resulting phase diagram in the $V \mhyphen t_1$ plane for $\delta=0.6$, $t_2 = 1.0$. Compared with the ED results, the mean-field calculations yield a considerable overestimation of the ordered phase, as expected. For a finite CDW order parameter, the chiral symmetry is broken, and the topology is lost. For $\delta_n =0$, on the other hand, the Hamiltonian reduces to the non-interacting case~\cite{Yin2018,Arouca2020}, and the boundaries between the topological phases are determined by computing the winding number through Eq.~\eqref{eq:winding_number}. 

\begin{figure}[t]
    \centering
    \includegraphics[scale = 0.5]{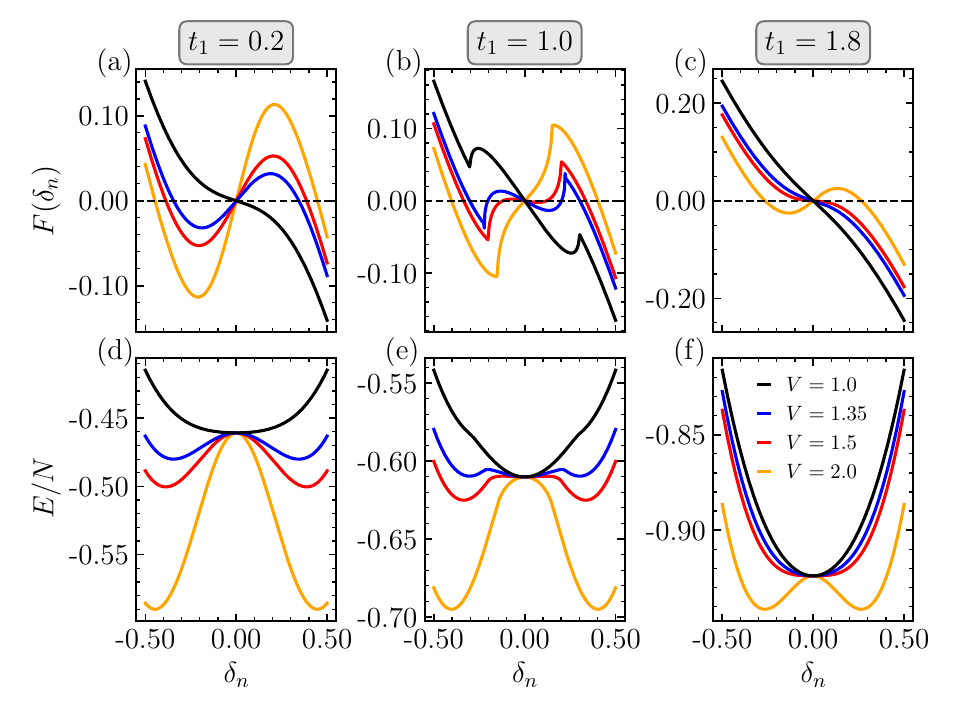}
    \caption{(a--c) Self-consistent function and (d--f) total energy as a function of $\dn$ for several values of $V$ with fixed $\delta = 0.6$ and $t_2 =1$.}
    \label{fig:gap_equation}
\end{figure}

The results for Eqs.~\eqref{Eq:mf_energy} and \eqref{Eq:self_consist_MF} reveal that the Topo-II-to-CDW transition is actually first order, while the Topo-I-to-CDW and Trivial-to-CDW remain of second order. This is summarized in Fig.~\ref{fig:gap_equation}, with $t_1 = [0.2,1.0,1.8]$, which are representative values of the Topo-I, Topo-II, and Trivial phases at small $V$.

\section{Spectral signatures of \texorpdfstring{$\mathcal{P}\mathcal{T}$}{PT} symmetry breaking}
\label{App:pt_symm_eigen}

\begin{figure}
    \centering
    \includegraphics[scale = 0.5]{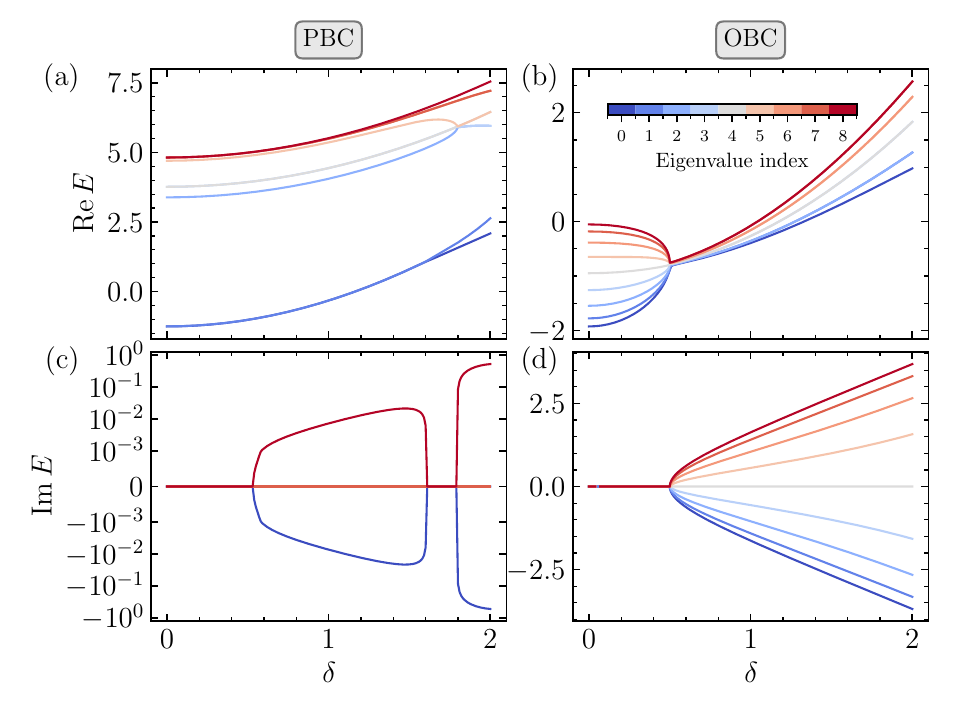}
    \caption{Real part of the eigenvalue spectrum for periodic (a) and open (b) boundary conditions, panels (c) and (d) show the corresponding imaginary part. All data for $N = 16$, $t_1 =0.5$, $t_2 =1$ and $V = 8$.}
    \label{fig:PT_broken_energies}
\end{figure}

The non-Hermitian SSH model can exhibit exceptional points as the hopping non-reciprocity is varied. At such points, eigenvalues coalesce, and the Hamiltonian becomes defective, leading to an ill-conditioned eigenbasis. In the non-interacting limit, this can produce a collapse of the single-particle spectrum for special parameter values, including $\delta=t_1$. In the interacting many-body problem, the spectrum is modified by interactions, and the location and structure of these exceptional features need not coincide with those of the single-particle case. Nevertheless, the approach to an exceptional line can still be detected through a strong clustering of low-lying many-body eigenvalues and the emergence of complex eigenenergies, which are signatures of $\mathcal{PT}$-symmetry breaking when the corresponding $\mathcal{PT}$ symmetry is present~\cite{El-Ganainy2018, Ashida2020}.

Figure~\ref{fig:PT_broken_energies} displays the real and imaginary parts of the low-lying many-body spectrum within the CDW phase, under PBC and OBC. While the real part of the spectrum retains a nearly degenerate ground state and a very small imaginary part for PBC, the spectrum of the OBC case is dramatically affected near the $\delta \to t_1$ EP. This spectral rearrangement is consistent with an enhanced low-energy response near the exceptional line and provides a possible origin for the enhanced charge correlations observed under OBC. It also underscores the qualitative difference between PBC and OBC in the non-Hermitian problem.

\section{Phase diagram under Open boundary conditions}
\label{App:OBC_phase_diag}
\begin{figure}
    \centering
    \includegraphics[scale = 0.5]{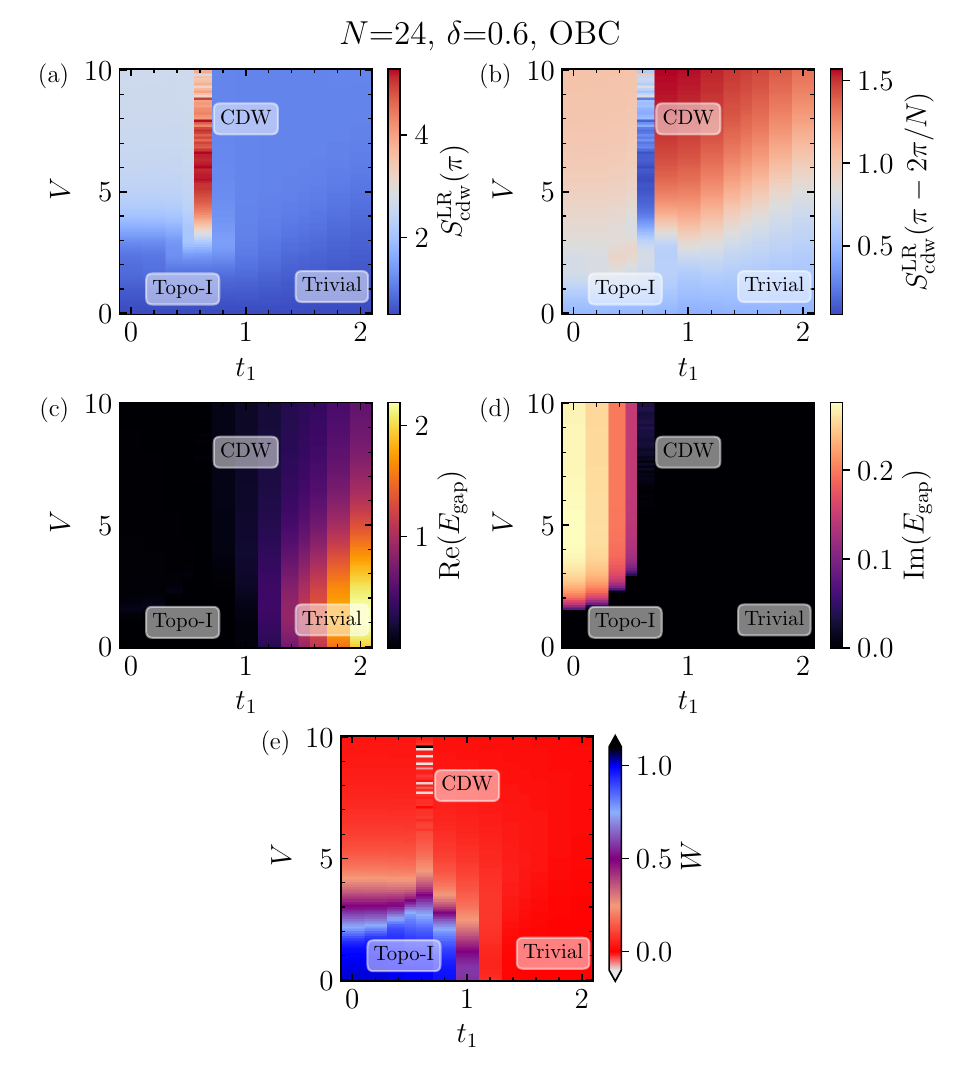}
    \caption{The charge structure factor for $q=\pi$ (a), and $q=\pi-2\pi/N$ (b), real (c) and imaginary (d) many-body energy gaps, as functions of $V$ and $t_1$. Data for $\delta=0.6$, $t_2 = 1.0$ and $N = 24$ under OBC. Panel (e) presents the topological marker. All data computed from left and right eigenvectors.
    }
    \label{fig:OBC_phase_diag}
\end{figure}
Under OBC, both the energy spectrum and the ground-state phase diagram differ from their counterparts under PBC. Figure~\ref{fig:OBC_phase_diag} summarizes these results. As discussed in the main text, one important difference in the OBC case is that the peak of the charge structure factor shifts from $q=\pi$ [Fig.~\ref{fig:OBC_phase_diag}(a)] to $q=\pi-2\pi/N$ [Fig.~\ref{fig:OBC_phase_diag}(b)] across the line $t_1=\delta$ within the CDW phase. Across this line, the many-body spectrum also changes its character, exhibiting a real energy gap for $t_1>\delta$ [Fig.~\ref{fig:OBC_phase_diag}(c)] and an imaginary energy gap for $t_1<\delta$ [Fig.~\ref{fig:OBC_phase_diag}(d)] inside the CDW phase. For weak interactions, the system exhibits only a gapless topological phase with $W \to 1$ and a gapped trivial phase with $W\to 0$, as shown in panel (e). These quantities allow us to characterize the full $t_1 \mhyphen V$ phase diagram.

\section{Comparison with the $\eta = \rm R$ case}
\label{App:left_right_comparison}

\begin{figure}
    \centering
    \includegraphics[scale = 0.5]{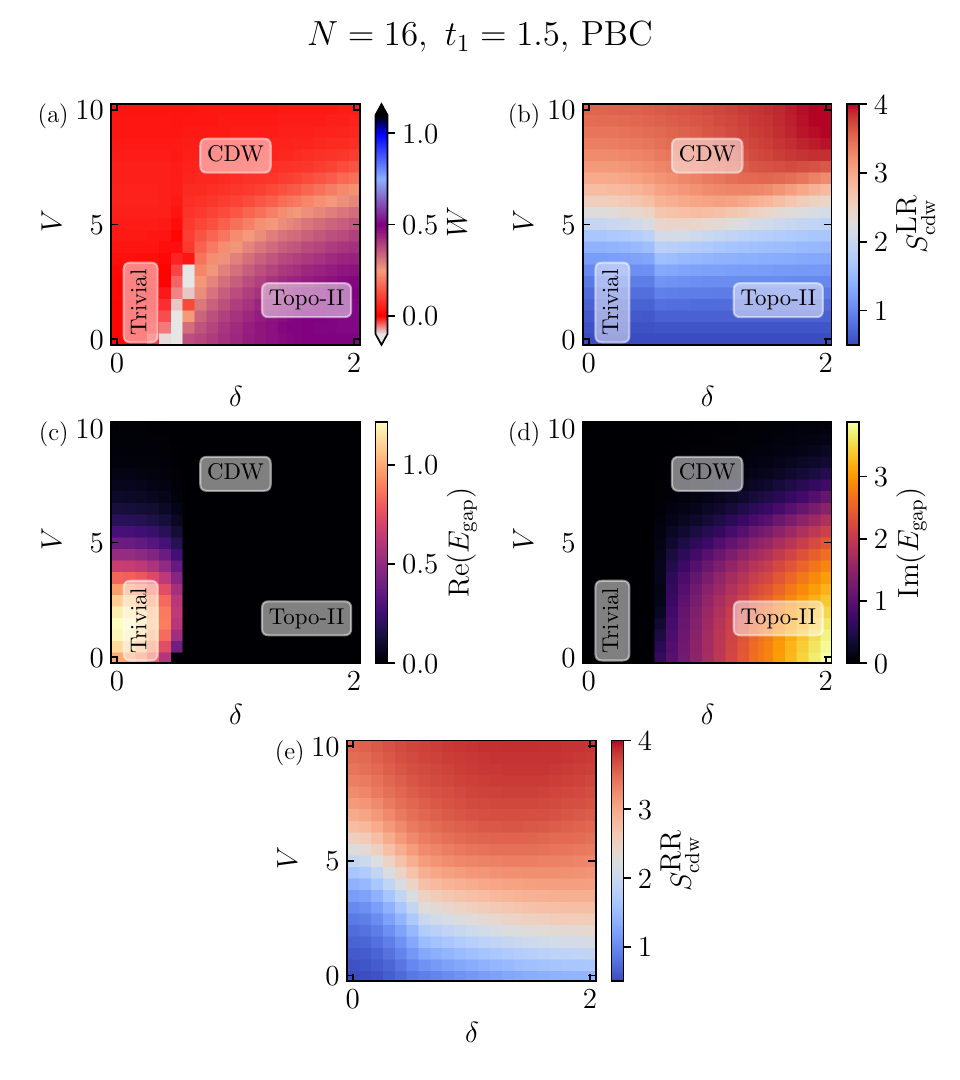}
    \caption{The topological maker (a), charge structure factor from left and right eigenvectors (b), real (c), and imaginary (d) many-body energy gaps. Data for $t_1=1.5$, $t_2 = 1$ and $N = 16$ under PBC. In panel (e), $S_{\rm cdw}$ is computed only with right eigenvectors $\eta={\rm R}$.}
    \label{fig:diagram_V_vs_delta_lr_rr}
\end{figure}

In the main text, all physical observables are computed using the biorthogonal left-right convention, $\eta={\rm L}$. This choice is natural for a non-Hermitian Hamiltonian, since expectation values are evaluated with the left and right eigenstates of the same eigenvalue. With this convention, we observe consistent agreement among the diagnostics summarized in Table~\ref{tab:phase_diagram} across the $V \mhyphen t_1$ phase diagram. As a matter of contrast, Fig.~\ref{fig:diagram_V_vs_delta_lr_rr} shows the corresponding results for the $V \mhyphen \delta$ phase diagram, where Fig.~\ref{fig:diagram_V_vs_delta_lr_rr}(a)-\ref{fig:diagram_V_vs_delta_lr_rr}(d) again demonstrate consistency in the identification of all phases. However, when the charge structure factor is computed using only the right eigenvectors, $S_{\rm cdw}^{\rm RR}$ [Fig.~\ref{fig:diagram_V_vs_delta_lr_rr}(e)] may incorrectly suggest a coexistence between the Topo-II and CDW phases. This is inconsistent with the chiral-symmetry-based characterization used in the main text: while neither the hopping terms nor the nearest-neighbor interaction breaks the chiral symmetry, the emergence of a CDW phase breaks the chiral symmetry in the thermodynamic limit, which in turn destroys the topological phase.

\begin{figure}
    \centering
    \includegraphics[scale = 0.5]{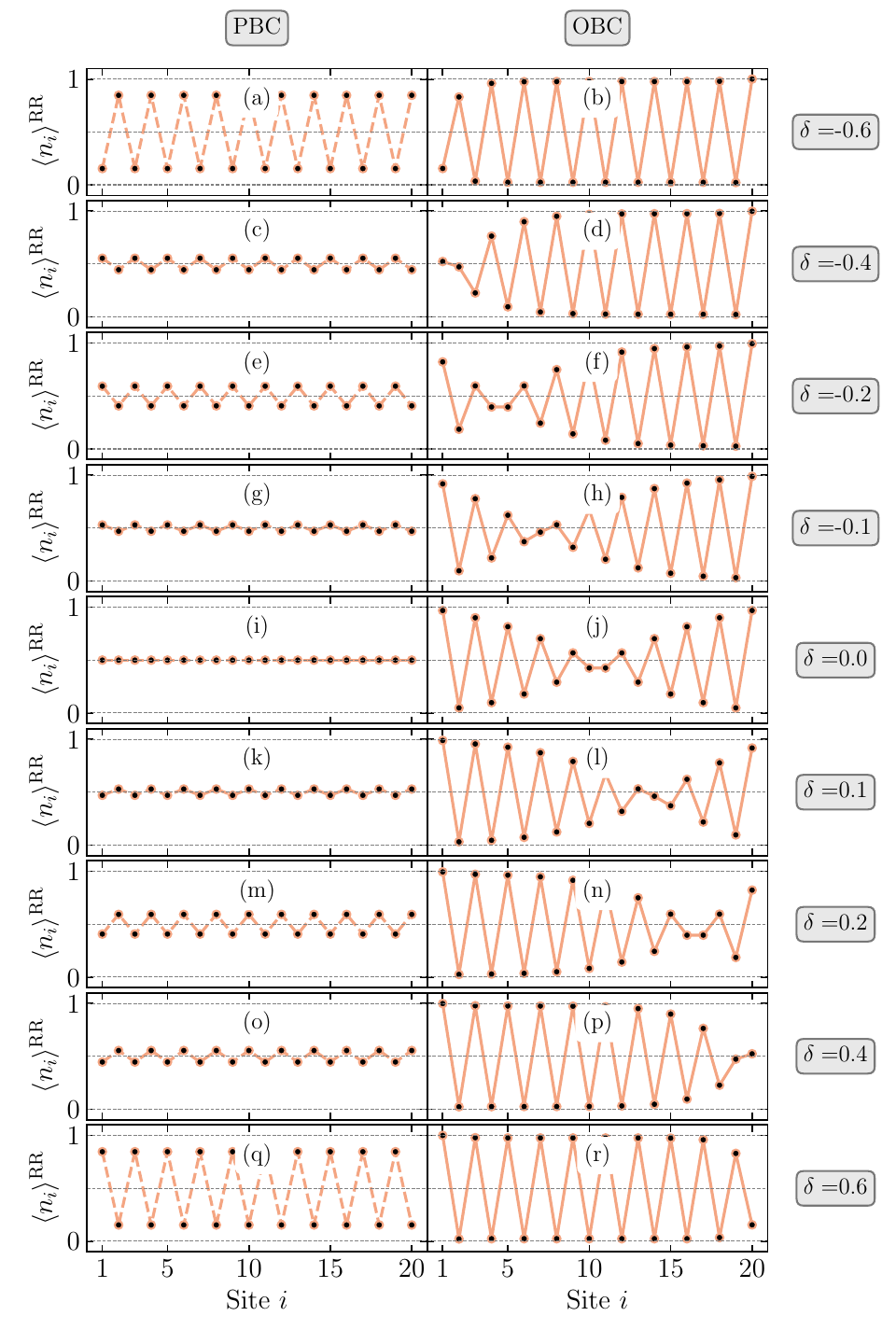}
    \caption{Real-space density profile for several values of $\delta$ under PBC and OBC, computed only with the right eigenvectors. Data for $N=20$, $t_1 =0.7$, $t_2=1$ and $V = 7$.}
    \label{fig:local_density_pbc_obc_rr}
\end{figure}

In addition, an even more problematic issue emerges when considering RR correlations. In finite systems under PBC, the average charge density at each site should remain uniform across the entire lattice. This is indeed the case when we compute the density using left and right eigenvectors, as in Eq.~\ref{eq:scdw_lr_rr} of the main text with $\eta = L$. Using $ \eta=R$ under PBC instead, produces staggered density patterns as shown in Fig.~\ref{fig:local_density_pbc_obc_rr}. This artifact is also observed in the enhanced staggered charge structure factor for $\eta = \rm R$ in Fig.~\ref{fig:diagram_V_vs_delta_lr_rr}(e), even inside the Topo-II phase. 

For OBC, there is also another major difference between the $\eta=\rm R$ and the $\eta=\rm L$ case. As discussed in Ref.~\cite{Zhong2025}, when one computes the local charge density using only the right eigenvectors, a ``kink'' is observed, and the charge density is no longer symmetric with respect to the center of the lattice. Here we also observe this for OBC; as shown in Fig.~\ref{fig:local_density_pbc_obc_rr}, the position of the kink is mostly controlled by the value of $\delta$. However, this feature seems to be an artifact emerging from the ``right'' eigenvectors, and as one computes the $\eta=\rm L$ case, there is no kink moving from the center sites for arbitrary $\delta$ values, and the density is symmetric with respect to the center of the lattice, apart from the vicinity of the EP, as discussed in terms of Fig.~\ref{fig:local_density_obc_lr} of the main text. 

\section{Topological phase transition under weak interactions}
\label{app:weak_coupling}

In the weak-coupling regime, interactions renormalize the energy spectrum and modify the topological phase boundaries without inducing charge order.  Figure~\ref{fig:weak_coupling_topo_transition} presents the results for $V=1$ under both PBC and OBC. In the Topo-I phase under PBC, the real part of the many-body gap is finite, while the imaginary part vanishes for sufficiently large system sizes at small $t_1$. As $t_1$ increases, the real part of the many-body gap [Fig.~\ref{fig:weak_coupling_topo_transition}(a)] vanishes in the interval $0.57 \lesssim t_1 \lesssim 1.52$, corresponding to the Topo-II phase. This region is instead characterized by a finite imaginary many-body gap [Fig.~\ref{fig:weak_coupling_topo_transition}(c)]. For even larger values of $t_1$, the gap becomes purely real again inside the trivial phase. Despite finite-size effects, the topological marker [Fig.~\ref{fig:weak_coupling_topo_transition}(e)] still captures the transitions between the Topo-I ($W \sim 1$), Topo-II ($W \sim 1/2$), and trivial ($W \sim 0$) phases.

\begin{figure}[H]
    \centering
    \includegraphics[scale = 0.5]{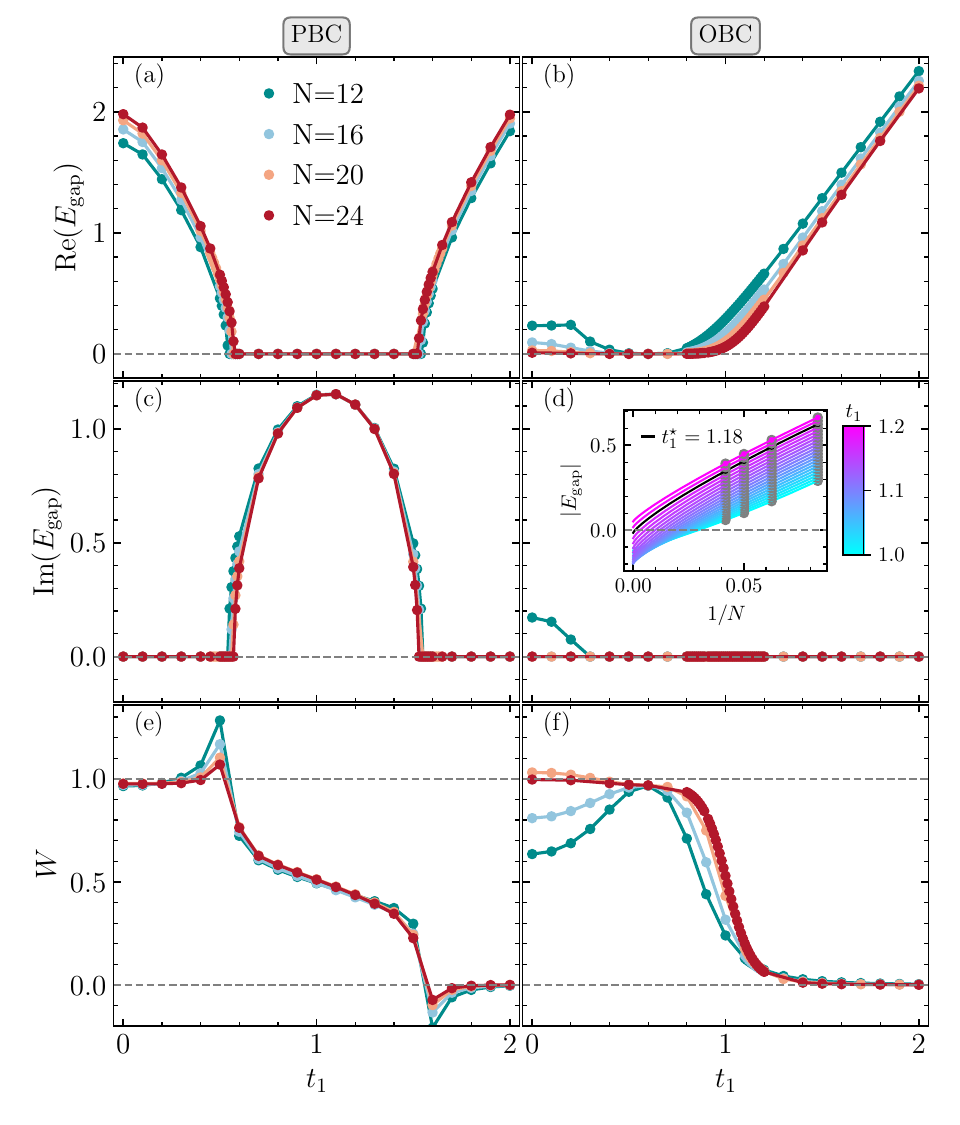}
    \caption{(a) Real and (c) imaginary parts of the many-body energy gap, together with (e) the topological marker, as functions of $t_1$ under PBC. Panels (b), (d), and (f) display the corresponding quantities under OBC. The inset in panel (d) shows the finite-size scaling of the many-body gap; the black line marks the critical value $t_1^\star = 1.18$, which separates the topological and trivial phases. Data are for $\delta = 0.6$, $t_2 = 1$, and $V = 1$.}
    \label{fig:weak_coupling_topo_transition}
\end{figure}

Under OBC, on the other hand, the gap closings and openings are less sharply resolved than in the PBC case, a direct consequence of the larger finite-size effects in the former. In this case, the imaginary part of the gap vanishes [Fig.~\ref{fig:weak_coupling_topo_transition}(d)], while the real part [Fig.~\ref{fig:weak_coupling_topo_transition}(b)] remains finite only inside the trivial phase. In this case, to estimate the critical point more accurately, we extrapolate the many-body gap to the thermodynamic limit using a power-law fit as $1/N \to 0$. This analysis yields the critical value $t_1^\star \sim 1.18$, indicated by the black line in the inset of Fig.~\ref{fig:weak_coupling_topo_transition}(d). As the system size increases, the topological marker becomes progressively sharper and consistently identifies the Topo-I ($W \to 1$) and trivial ($W \to 0$) phases, as shown in Fig.~\ref{fig:weak_coupling_topo_transition}(f). This analysis further allows us to determine the critical points along the horizontal cut of the phase diagrams shown in Fig.~\ref{fig:lattice} of the main text.

\end{twocolumngrid}



\end{document}